\documentclass[11pt,a4paper]{article}

\usepackage{jheppub}
\usepackage{amsmath,amssymb}
\usepackage[utf8]{inputenc}
\usepackage{color}
\usepackage{graphicx}
\usepackage{verbatim}
\usepackage{braket}
\usepackage{slashed}
\allowdisplaybreaks[1]

\newcommand{\KD}{K\"{a}hler-Dirac }

\def\psib{\overline{\psi}}
\title{Lorentzian K\"{a}hler-Dirac fermions}

\author[a]{Simon Catterall}
\author[a]{and Jay Hubisz}

\affiliation[a]{Department of Physics, Syracuse University, Syracuse, NY 13244, USA}

\emailAdd{smcatter@syr.edu}
\emailAdd{jhubisz@syr.edu}

\abstract{
We examine the formulation of \KD fermions on spacetimes with Lorentz signature. In practice we focus
on Minkowski spacetime since most of the difficulties that are encountered
are visible even when the spacetime is flat. We show that the theory, when interpreted as a Lorentz
invariant theory of
forms or antisymmetric tensor fields, is non-unitary. We show that
unitarity can be restored provided one
adopts a modified inner product on the Hilbert space. This modified inner product requires the insertion of an 
operator $J$ that anti-commutes with certain modes of the \KD field in such a way as to guarantee all states
have positive norm. We give an explicit (and local) form for $J$ and show that 
while it commutes with
the Hamiltonian, it is incompatible with the Lorentz transformation properties of the tensor fields.  In flat space, the unitarized formulation is equivalent to 4 Dirac fermions.
}

\begin{document}

\maketitle

\section{Introduction}

At essentially the same time that Dirac published his famous equation for the electron
there was another proposal from Landau and Ivanenko in which the wavefunction of the electron
was represented as a set of antisymmetric tensor fields \cite{IvanenkoLandau:1928}. While this
tensor representation had the advantage of avoiding the use of spinors and gamma matrices it appeared to
describe not one but four degenerate electrons which was presumably seen as a deficiency
as compared with the Dirac equation. Indeed, after the success of the Dirac theory Landau's proposal
was largely forgotten
until it was rediscovered by K\"{a}hler decades later where it was generalized to curved space \cite{Kahler:1962}. This so-called \KD equation received
some attention in the early days of lattice gauge theory \cite{Graf:1978,BecherJoos:1982,BennTucker:1982,Goeckeler:1984} but was again
discarded after it was realized that its discrete form in flat space was equivalent to
staggered fermions \cite{BanksDothanHorn:1982, Susskind:1977}. \KD fermions
resurfaced once again in the construction of supersymmetric lattice theories - see the review \cite{Catterall:2009it} and references therein
and most recently
in the study of lattice anomalies and symmetric mass generation
\cite{Ayyar:2014eup,Ayyar:2016lxq,Slagle:2014kla,Butt:2018nkn,Catterall:2021anomaly,You:2014oaa,You:2017ltx,You:2017ofi,Catterall:2022jky,Maiti:2025smg,Maiti:2026phase,Wang:2022ucy}. 

One of the more remarkable results emerging from this recent work was the discovery
that (Euclidean) \KD fermions exhibit a new type of gravitational anomaly which can be computed exactly
in the discretized theory. This anomaly breaks a certain global $U(1)$ symmetry down to $Z_4$ for
manifolds with the topology of a sphere
\cite{Catterall:2018lkj}. Cancellation of a mod 2 't Hooft anomaly for this remaining discrete
symmetry then produces a series of theories that allow for symmetric mass 
generation~\footnote{In fact, if one decomposes the
\KD field into spinors, the number of
fermions needed to cancel the mod 2 \KD anomaly is equal to the number of
fermions needed to gap boundary states in topological superconductors without breaking symmetries. Equivalently it
reproduces the number of Majorana fermions needed to cancel certain Dai-Freed anomalies in these theories\cite{Guo:2023smg,GarciaEtxebarria:2018ajm,Catterall:2022jky,Butt:2021brl}.}.
In four dimensions the simplest anomaly free theory of massless \KD fermions can be decomposed in flat space into spinors and corresponds
to the Pati-Salam GUT \cite{Pati:1974yy,Catterall:2022jkn}~\footnote{Strictly it involves four copies of a reduced \KD field which carries half the degrees of freedom of the usual \KD field.}. This observation suggests that discrete \KD theories
might be relevant to the problem of constructing chiral lattice gauge theories. The \KD theory evades the usual
Nielsen-Ninomiya theorem by focusing not on the usual chiral symmetry but rather on a variant compatible with the tensor formulation that can be implemented as an onsite lattice symmetry without occasioning fermion doubling.

Unfortunately, almost the studies of \KD fermions to date have been done in Euclidean space. In this paper we
examine the theory in Lorentz signature. The starting point of our work has some overlap with recent work
by Boyle et al. \cite{Boyle:2025cpt} although our final conclusions differ.
One unsurprisingly encounters issues:  the numerous advantages of the K\"ahler-Dirac formulation are overshadowed by obvious conflict with the spin-statistics theorem.  As the fields are formulated as tensors, they are not in a spinorial representation of the Lorentz group.  This obstructs their formulation as a unitary, local, and Lorentz invariant theory for fermions.  

This difference is explicitly manifest in a spacetime with curvature, where the equation of motion and spectrum deviate from the Dirac equation in the same background.  We will also see that
the Wick rotation of the Lorentzian KD theory into Euclidean signature gives rise to some ``wrong sign" kinetic terms leading to a violation of reflection positivity.

Our paper focuses on a resolution to the issue of unitarity where the Hilbert space norm is modified in such a way as to give all states positive norm. We will see that this modification 
necessarily breaks the Lorentz transformation properties of the forms, and makes the theory indistinguishable (in flat space) from 4 copies of an ordinary Dirac fermion.  This then trivially renders the unitarized theory
compatible with the spin statistics theorem. 

\section{Matrix representation of a \KD field}
The massless K\"{a}hler-Dirac action is given by
\begin{equation}
    S=\int d^4x\,\sqrt{g}\left[\overline{\Phi}\,iK\,\Phi\right]
\end{equation}
where $K=(d-d^\dagger)$ is the K\"{a}hler operator, $d$ is the exterior derivative and the K\"{a}hler field $\Phi$ is a collection of p-forms
$\Phi=(\phi,\phi_\mu,\phi_{\mu\nu},\ldots)$ with $\overline{\Phi}=(\phi^*,\phi^{*\mu},\phi^{*\mu\nu},\dots)$ corresponding to the set of all p-forms in $D$ dimensions \cite{BanksDothanHorn:1982}.
The bracket notation indicates an inner product over forms:
\begin{equation}
    \left[A, B\right]=\sum_{p=0}^D A_{\mu_1\ldots\mu_p} B^{\mu_1\ldots\mu_p}
    \label{eq:innerprod}
\end{equation}
In general we can map such a theory of forms into a theory of matrix valued fields $\Psi$ by using the forms as
coefficients in an expansion over the Clifford algebra.
In practice in this paper we will limit ourselves to flat four dimensional spacetime where this
expansion is given explicitly as
\begin{equation}
    \Psi=\phi I+\phi_\mu\gamma^\mu+\phi_{\mu\nu}\gamma^\mu\gamma^\nu+\phi_{\mu\nu\lambda}\gamma^\mu\gamma^\nu\gamma^\lambda+\phi_{0123}\gamma^0\gamma^1\gamma^2\gamma^3
    \label{eq:matfield}
\end{equation}
Using this representation allows us to write the original action in terms of matrices
\begin{equation}S=\frac{1}{4}\int d^4x\,{\rm Tr}\left[\gamma^0\Psi^\dagger\gamma^0\left(i\gamma^\mu\partial_\mu \right)\Psi\right]\label{matrixaction}\end{equation}
It is immediately clear that the equations of
motion of the matrix theory in flat spacetime  describe four degenerate Dirac fermions.

However, notice that in going to the matrix representation we have replaced the complex conjugate \KD field $\bar{\Phi}$ by $\gamma^0\Psi^\dagger\gamma^0$.  This is the appropriate Dirac adjoint of the matrix field $\Psi$.  The factors of $\gamma^0$ can be seen to arise from the fact that the spatial $\gamma$-matrices in Lorentz signature are not hermitian.  Rather, we have  $(\gamma^\mu)^\dagger = \gamma^0 \gamma^\mu \gamma^0$.  The $\gamma^0$ factors are required in order to consistently replace the inner product in Eq.~(\ref{eq:innerprod}) with the trace.  Additionally, the matrix field (\ref{eq:matfield}) rotates from both the left and right under Lorentz transformations, as we show below.  Therefore two $\gamma^0$ factors are required for the daggered field to have the correct rotation properties under multiplication by the (non-unitary) Lorentz tranformation matrices.

In flat spacetime, if we interpret the action as that of four flavors of Dirac fermion, the $\gamma^0$ on the right of $\Psi^\dagger$ is there to
ensure that the theory is Lorentz invariant under the usual transformation of spinors in Minkowski spacetime.  The additional factor of $\gamma^0$ means that the degenerate flavors are invariant not under $SU(4)$ 
flavor transformations as one might naively expect, but rather $\Psi\to \Psi F^{-1}$ satisfying
$F^{-1}\gamma^0(F^{-1})^\dagger=\gamma^0$ corresponding to the non-compact flavor group $U(2,2)$. 

The transformation properties of the matrix field follow from the invariance of
the matrix theory under the diagonal symmetry arising when a $SO(3,1)$ subgroup of $U(2,2)$ is locked to
the Lorentz group corresponding to the Lorentz transformation 
\begin{equation}
    \Psi\to L\Psi L^{-1}
    \label{lorentz}
\end{equation}
Under this diagonal subgroup the transformed $\Psi$ can be
written
\begin{align}
    L\Psi L^{-1}&=\phi I+\Lambda^{\mu}_{\nu}\gamma^\nu \phi_\mu+ \Lambda_\sigma^\mu\Lambda_\tau^\nu\gamma^\sigma\gamma^\tau\phi_{\mu\nu}+\ldots\nonumber\\
    &=\phi I+\left[\Lambda_\nu^\mu \phi_\mu\right]\gamma^\nu+\left[\Lambda_\sigma^\mu\Lambda_\tau^\nu\phi_{\mu\nu}\right]\gamma^{\sigma}\gamma^{\tau}+\ldots
\end{align}
Thus the equivalence between the matrix theory and the \KD form theory
is only possible if the field $\Psi$ transforms as in eqn.~\ref{lorentz}.

Let us now examine the discrete symmetries of the massless theory. Again, to retain the p-form
interpretation, these transformations must have equal left and right actions on $\Psi$. If this
were not the case then a mixing of the p-forms would take place and the assignment of the discrete symmetries would not be compatible with the Lorentz transformation properties of the matrix field.

Consider first time reversal $\cal T$. In chiral or Dirac basis this takes the form
\begin{align}
    \Psi(\vec{x},t)&\to \gamma^1\gamma^3\Psi(\vec{x},-t) \gamma^3\gamma^1\nonumber\\
    i&\to -i\nonumber\\
    t&\to -t
\end{align}
It is straightforward to show that this is an invariance of the action.
Similarly, a parity transformation $\cal P$ given by
\begin{align}
    \Psi(\vec{x},t)&\to \gamma^0\Psi(-\vec{x},t)\gamma^0\nonumber\\
    \vec{x}&\to -\vec{x} 
\end{align}
is also easily seen to be symmetry of the action.
The natural choice for charge conjugation $\cal C$
corresponds to the transformation
\begin{equation}
    \Psi\to i\gamma^2  \Psi^*i\gamma^2 
\end{equation}
In this case
\begin{align}
S^\prime &=\int d^4x\,{\rm Tr}
\left[\gamma^0 (i\gamma^2\Psi^*i\gamma^2)^\dagger\gamma^0(i\gamma^\mu\partial_\mu)i\gamma^2 \Psi^*i\gamma^2 \right]\nonumber\\
&=\int d^4x\,{\rm Tr}
\left[  \gamma^0 \Psi^T \gamma^0  i\gamma^2 (i \gamma^\mu \partial_\mu) i\gamma^2  \Psi^* \right]\nonumber\\
&=\int d^4x\,{\rm Tr}
\left[  \gamma^0 \Psi^\dagger \gamma^0 \gamma^0i\gamma^2 ( i (\gamma^\mu)^T \partial_\mu ) i\gamma^2\gamma^0 \Psi \right]\nonumber\\
&=\int d^4x\,{\rm Tr}
\left[\gamma^0\Psi^\dagger\gamma^0(-i\gamma^\mu\partial_\mu)\Psi\right],
\end{align}
where we have used $C(\gamma^\mu)^T C^{-1}=-\gamma^\mu$ with $C=i\gamma^2\gamma^0$.
It is thus not a symmetry of the kinetic term. The simplest fix is to change the transformation to 
\begin{equation}
    \Psi\to \gamma^5i\gamma^2\Psi^*i\gamma^2\gamma^5
\end{equation}
This has the added merit that its net action on $\Psi$ is to simply complex conjugate the
p-forms $\phi_{\mu_1\cdots\mu_p}$.
There are two possible mass terms that are compatible with the p-form interpretation of the theory:
\begin{align}
&m \mathrm{Tr} \left[ \gamma^0\Psi^\dagger\gamma^0\Psi \right] \nonumber\\
\mathrm{~~or~~} &m \mathrm{Tr} \left[ \gamma^0\Psi^\dagger\gamma^0\gamma^5\Psi\gamma^5  \right]
\end{align}
These are $\cal T$ and $\cal P$ invariant, but both of these possible mass terms are odd under $\cal C$.  We conclude that the theory, interpreted as a collection of forms, is ${\cal C}{\cal P}{\cal T}$ invariant only if the mass term is set to zero. This, at first sight rather surprising result, can be explained by noting
that the kinetic operator has another axial-like symmetry corresponding to
\begin{equation}
    \Psi\to e^{i\alpha\gamma^5\otimes\gamma^5}\Psi
\end{equation}
where the notation indicates that $\gamma^5$ acts on left and right of the matrix fermion.
This symmetry alone is sufficient to forbid mass terms and for $\alpha=\frac{\pi}{2}$
corresponds to a left and right action of $\gamma^5$ on $\Psi$. Thus the natural 
charge conjugation symmetry is intertwined with a discrete axial symmetry and hence prohibits mass terms. Notice that $\cal C$ does allow for four fermion or gauge interactions.

Using $\cal C$ we can define the analog of a Majorana condition for \KD fields:
\begin{equation}
    \Psi=\Psi^c=\gamma^5 i\gamma^2\Psi^* i\gamma^2\gamma^5.
\end{equation}
This is just a reality condition on the forms, and it implies that
\begin{align}
\gamma^0\Psi^\dagger\gamma^0&=\gamma^0(\gamma^5i\gamma^2 \Psi^T i\gamma^2\gamma^5)\gamma^0\nonumber\\
    &=\gamma^5(C^{-1}\Psi^T C)\gamma^5.
\end{align}
The massless Majorana \KD action can then be written
\begin{equation}
    S_{\rm M}=\int d^4x\,{\rm Tr}\left[\gamma^5(C^{-1}\Psi^TC)\gamma^5 i\gamma^\mu\partial_\mu\Psi\right]
\end{equation}
In terms of forms this is
\begin{equation}
    S_M=\int d^4x\,\left[\Phi\, iK\, \Phi\right].
\end{equation}
This action describes a system with one half the number of degrees of freedom and describes four Majorana
fermions if the \KD field is interpreted as spinors. The Majorana condition breaks the original
$U(2,2)$ flavor symmetry down to $SO(3,1)$ corresponding to the generators that
commute with $\gamma^5\gamma^0\gamma^2$. This reduced flavor symmetry
is then locked with the Lorentz symmetry to give the
interpretation in terms of real-valued form fields.

\section{Canonical quantization and the emergence of negative norm states}
While the \KD theory is Lorentz invariant it does suffer from a major
problem - the presence of the extra left acting $\gamma^0$ in the matrix action in eqn.~\ref{matrixaction} leads to a violation of unitarity. One can see this clearly after
decomposing $\Psi$ into left eigenstates of $\gamma^0$. 
\begin{align}
    \Psi_\pm=\Psi\frac{1}{2}(1\pm\gamma^0)
\end{align}
In terms of the projected fields one can easily write
the action as
\begin{equation}
    S=\int d^4x\,{\rm Tr}\,[\Psi^\dagger_+ i\partial_t\Psi_+-\Psi^\dagger_+i\alpha^i\partial_i\Psi_+]-
    \int d^4x\,{\rm Tr}\,[\Psi^\dagger_-i\partial_t\Psi_--\Psi^\dagger_-i\alpha^i\partial_i\Psi_-]
\end{equation}
where $\alpha^i=\gamma^0\gamma^i$. The corresponding Hamiltonian is
\begin{equation}
    H=\int d^3x\,{\rm Tr}\left(\Psi_+^\dagger i\alpha^i\partial_i\Psi_+\right)-\int d^3x\,{\rm Tr}\left(\Psi_-^\dagger i\alpha^i\partial_i\Psi_-\right)
    \label{Hamil}
    \end{equation}
and the canonical equal-time anti-commutators take the form
\begin{align}
    \{\Psi_+(x),\Psi^\dagger_+(y)\}&=\delta^3(x-y)\nonumber\\
    \{\Psi_-(x),\Psi^\dagger_-(y)\}&=-\delta^3(x-y)
    \label{anti}
\end{align}
In a periodic box one finds the corresponding anti-commutators in momentum space:
\begin{align}\{\tilde{\Psi}_+(k),\tilde{\Psi}^\dagger_+(q)\}&=\delta_{kq}\nonumber\\
\{\tilde{\Psi}_-(k),\tilde{\Psi}^\dagger_-(q)\}&=-\delta_{kq}
\label{anti2}
\end{align}
The presence of the extra minus signs for the $\Psi_-$ modes
in the RHS of eqn.~\ref{anti} and eqn.~\ref{anti2} leads to negative norm states and a non-unitary
theory if one follows standard quantization procedures. One can see this explicitly by computing the norm of a single particle state
\begin{equation}
 \bra{0}\tilde{\Psi}_-(k) \tilde{\Psi}_-^\dagger(k)\ket{0}=\bra{0}(-\tilde{\Psi}^\dagger_-(k)\tilde{\Psi}_-(k)-1)\ket{0}=-\braket{0|0}
\end{equation}
Indeed the Hamiltonian for the $\Psi_-$ modes as written
in eqn.\ref{Hamil} also carries a wrong sign. This problem was also noted in \cite{Boyle:2025cpt}.

This failure of unitarity is reflected in the fact that the Euclidean theory 
is not reflection positive.  Reflection positivity requires that $<\Theta F F>$ is positive semi-definite for $F$ some
function of the fields and $\Theta$ a reflection operation.
In the \KD case this is given by
\begin{equation}    
\Psi(x_0,\vec{x})\stackrel{\Theta}{\to} \gamma^0\Psi(-x_0,\vec{x})\gamma^0
\end{equation}
where again the form interpretation forces a left-right action of $\gamma^0$. This implies
\begin{align}
    \Psi_+(x_0,\vec{x})&\to \Psi_+(-x_0,\vec{x})\gamma^0\nonumber\\
    \Psi_-(x_0,\vec{x})&\to -\Psi_-(-x_0,\vec{x})\gamma^0
\end{align}
Since $\Psi_+$ behaves like two vanilla Dirac fermions under $\Theta$ the proof of reflection positivity goes through as usual. However, the extra minus sign in the $\Psi_-$ transformation
translates into a violation of this condition.

\section{A toy model: the J-modified inner product}
To try to understand how to proceed let us consider a simple toy model where
a minus sign also appears in the canonical anti-commutator in the same way.
We start by considering the unitary case where the wrong signs are not present and then graduate to the non-unitary case.
\subsection{The unitary case}
Consider 
a single fermionic degree of freedom corresponding to the operators $\Psi$ and $\Psi^\dagger$ with Lagrangian 
\begin{equation} L=\Psi^\dagger i\frac{\partial}{\partial t}\Psi-\Psi^\dagger\Psi\end{equation}.
The canonical anti-commutators are
\begin{equation}
    \Psi^2=(\Psi^\dagger)^2=0\quad{\rm and}\quad \{\Psi,\Psi^\dagger\}=1\label{simple}.
\end{equation}
Clearly this is a 2 state system with
\begin{equation}
    \Psi^\dagger\ket{0}=\ket{1}\quad {\rm and}\quad \Psi\ket{1}=\ket{0}.
\end{equation}
Since 
\begin{equation}
(\Psi^\dagger\Psi)^2=\Psi^\dagger\Psi(-\Psi\Psi^\dagger+1)=\Psi^\dagger\Psi,
\end{equation}
the eigenvalues of $H=\Psi^\dagger\Psi$ are $0$ and $1$.
Clearly
\[\braket{1|1}=\bra{0}\Psi\Psi^\dagger\ket{0}=\bra{0}(-\Psi^\dagger\Psi+1)\ket{0}=\braket{0|0}\]
and all states have positive norm. In appendix B we show how unitarity is crucial to
the construction of a path integral for this system.

\subsection{The non-unitary case}
Suppose that the Lagrangian were instead given by
\[L=-\left(\Psi^\dagger i\frac{\partial}{\partial t}\Psi-\Psi^\dagger\Psi\right)\]
This implies that the canonical anti-commutation relation reads
\begin{equation}
    \{\Psi,\Psi^\dagger\}=-1
\end{equation}
Clearly $\Psi^\dagger\Psi$ will now have eigenvalues $0$ and $-1$ although the Hamiltonian
$H=-\Psi^\dagger\Psi$ will still have a positive spectrum. More importantly there
will now be a negative norm state 
\begin{equation}
\braket{1|1}=\braket{0|\Psi\Psi^\dagger| 0}=\bra{0}(-\Psi^\dagger\Psi-1)\ket{0}=-\braket{0|0}
\end{equation}
However,
we can remove the negative norm states by defining a new norm of the form $\braket{1|1}_J=\bra{1}J\ket{1}$ 
where we insert an operator $J$ that anti-commutes
with $\Psi$ and $\Psi^\dagger$. This means that
\[\braket{1|J|1}=\bra{0}\Psi J\Psi^\dagger\ket{0}=-\bra{0}\Psi\Psi^\dagger J\ket{0}=-\bra{0}(-\Psi^\dagger\Psi-1)J\ket{0}=\braket{0|J|0}\]
We will additionally require that $J$ is both hermitian and unitary so $J^2=1$.
This implies that
\begin{align}
J\Psi^\dagger J&=-\Psi^\dagger\nonumber\\
J\Psi J&=-\Psi
\end{align}
Furthermore, $J$ clearly commutes with the
Hamiltonian which guarantees that this positivity property is time independent.\footnote{One can generalize this statement to a system whose Hamiltonian is merely pseudo-hermitian $H^\dagger=JHJ$ - see appendix A.}
The following operator gives an explicit representation of $J$ for this toy model:
\begin{equation}
J=e^{i\pi\Psi^\dagger\Psi}=1+2\Psi^\dagger\Psi
\end{equation} 
where we have used the fact that $\Psi^\dagger\Psi$ has eigenvalues zero and minus one. This is similar
to the construction given in \cite{LeClair:2007jhep} for symplectic fermions.

The use of a J-norm  is also crucial to formulate a Euclidean path integral that could evaluate the
partition function of the model. The key observation is that such a construction requires a well
defined resolution of the identity which is absent for a non-unitary theory. In appendix B we
show how the introduction of a J-modified norm solves this problem and allows for the construction
of a Euclidean path integral for the J-deformed toy model along the same lines as given in
\cite{Ryu:2023prl}.

In summary, by introducing a unitary operator $J$ that
defines a new positive definite norm on the Hilbert space, the theory can be rendered manifestly unitary.

\section{A unitary theory of \KD fields}
\subsection{The matrix formulation}
As with the toy model, the appearance of negative norm
states in the \KD theory
can be avoided if one is able to find an operator $J$ with similar properties:
$J^2=1$ and 
\begin{align}
    J\tilde{\Psi}_-(k)J&=-\tilde{\Psi}_-(k)\nonumber\\
    J\tilde{\Psi}_-^\dagger(k)J&=-\tilde{\Psi}_-^\dagger(k)\nonumber\\
    J\tilde{\Psi}_+(k)J&=\tilde{\Psi}_+(k)\nonumber\\
    J\tilde{\Psi}_+^\dagger(k)J&=\tilde{\Psi}_+^\dagger(k).
\end{align}
The following operator does the job
\[J=e^{i\pi\int d^3k\,\tilde{\Psi}_-^\dagger(k)\tilde{\Psi}_-(k)}=\prod_k\left[1+2\tilde{\Psi}_-^\dagger(k)\tilde{\Psi}_-(k)\right]. \]
The same anticommutation properties with $J$ also hold for the field operators in momentum space.
Clearly insertion of this $J$ into the inner product then makes the norm of the state $\Psi_-^\dagger(k)\ket{0}$ positive. This argument can be generalized to multiparticle states of the form
\begin{equation}
   \prod_{i=1}^N \tilde{\Psi}_-^\dagger(k_i)\ket{0}.
\end{equation}
The J-norm of such a state is given by
\begin{equation}
    \bra{0}\prod_{j=N}^1 \tilde{\Psi}_-(k_j)\cdot J \cdot\prod_{i=1}^N \tilde{\Psi}^\dagger_-(k_i)\ket{0}.
\end{equation}
Anticommuting each $\tilde{\Psi}_-(k_m)$ to the right through $J$ produces a minus sign which then compounds with another
minus sign arising from the wrong sign anticommutator as one then pushes $\tilde{\Psi}_-(k_m)$ to the right of $\tilde{\Psi}_-^\dagger(k_m)$. The net result is a positive contribution from each term in the product. Adding $\tilde{\Psi}_+$ operators does not change this since they commute with $J$ and have the right sign anticommutation relation.
Thus the J-modified inner product leads to a positive norm Hilbert space. Notice that $J$ commutes with the Hamiltonian and is hence conserved by the dynamics. Indeed it can be written as
$\left(-1\right)^{N^{(-)}}$ where $N^{(-)}=\int d^3k\,\tilde{\Psi}_-^\dagger(k)\tilde{\Psi}_-(k)$ is the number of negative mode fields.

For this procedure to work as described, $\Psi_+$ must not mix with $\Psi_-$ under interactions of the theory.  Otherwise, $J$ will not commute with $H$. Also, symmetries of the action may not match symmetries of the matrix elements, since they may not commute with $J$.  Since boosts of the antisymmetric forms mix $\Psi_+$ and $\Psi_-$, they cannot be a symmetry
of the theory. Indeed, it is easy to see that
matrix elements defined by the J-deformed inner product are
not invariant under the full Lorentz symmetry but only its compact subgroup - the group of
matrices $R$ that generate the rotation group. Explicitly, 
under rotations $\Psi^\prime= R\Psi R^\dagger$ the key properties of the operator $J$ are retained:
\begin{equation}
    (J^\prime)^2=R(1+2\Psi_-^\dagger\Psi_-)R^\dagger R(1+2\Psi_-^\dagger\Psi_-)R^\dagger=1
\end{equation}
and 
\begin{equation}J^\prime\Psi^\prime_-=RJR^\dagger R\Psi_- R^\dagger=RJ\Psi_- R^\dagger=-\Psi_-^\prime J^\prime\end{equation}
Thus the price of unitarizing the \KD theory is that is no longer Lorentz invariant under both left and right action on the matrix fields. Of course the Hamiltonian is not manifestly Lorentz invariant but we learn from this argument that the J-deformed theory must necessarily explicitly break Lorentz invariance.  

One might ask what theory is obtained by constructing a partition function for this J-deformed
theory and interpreting it as a Euclidean path integral.
As discussed in appendix B this path integral will involve an insertion of $J$:
\begin{align}
    Z&={\rm Tr}\left(Je^{-\beta H}\right)={\rm Tr}\left(\left(-1\right)^{N_-} e^{-\beta H}\right)
\end{align}
This Euclidean theory, which is now reflection positive, is clearly not invariant under the analog of boosts in Euclidean space since $J$ (or equivalently 
$N_-$) is not Lorentz invariant. In $D$ Euclidean dimensions it is invariant only under $SO(D-1)$ rotations.

\subsection{Unitarity in the tensor formulation}
It is instructive to re-examine these unitarity issues from the point of view of
the tensor formulation of the theory. We start by decomposing the matrix fermion
in terms of eigenstates of the twisted chiral symmetry operator $\gamma^5\otimes \gamma^5$ as $\Psi=\Psi_A+\Psi_B$
where 
\begin{align}
\Psi_A=\frac{1}{2}(\Psi+\gamma^5\Psi\gamma^5)\nonumber\\
\Psi_B=\frac{1}{2}(\Psi-\gamma^5\Psi\gamma^5)
\end{align}
Substituting this expansion into the matrix action yields
\begin{equation}
    S=\int d^4x\,{\rm Tr}\left[\Psi_A^\dagger\left(i\partial_t+i\alpha_i\partial_i\right)\Psi_B\beta\right]+
    \left[\Psi_B^\dagger\left(i\partial_t+i\alpha_i\partial_i\right)\Psi_A\beta\right]
\end{equation}
where the matrix fields can be written explicitly as
\begin{align}
    \Psi_A&=aI+a_i\alpha^i+a_{ij}\alpha_i\alpha_j+a_{123}\alpha_1\alpha_2\alpha_3=\sum_{\vec{m}}a_{\vec{m}}\alpha_{\vec{m}}\nonumber\\
    \Psi_B&=b\beta +b_i\beta\alpha_i+b_{ij}\beta\alpha_i\alpha_j+b_{123}\beta\alpha_1\alpha_2\alpha_3=\sum_{\vec{m}}b_{\vec{m}}(\beta\alpha_{\vec{m}})
\end{align}
with $\alpha_{\vec{m}}=\alpha^{m_1}\alpha^{m_2}\alpha^{m_3}$ with $m_i=0,1$ ($\beta\equiv\gamma^0$). 
The two matrix bases $\{\alpha_{\vec{m}}\}$ and $\{\beta\alpha_{\vec{m}}\}$ transform as spatial vectors under rotations:
\begin{align}
    R\alpha_i R^\dagger&=r_{ij}\alpha_j\nonumber\\
    R(\beta\alpha_i)R^\dagger&=r_{ij}(\beta\alpha_j)
\end{align}
with $R=e^{\theta\gamma_i\gamma_j}$ and $r_{ij}$ a rotation matrix. This allows us to interpret $A\equiv (a,a_i,a_{ij},a_{123})$ and
$B\equiv (b,b_i,b_{ij},b_{123})$ as two independent {\it spatial} K\"{a}hler-Dirac fields. 
The action becomes
\begin{align}
    S=&\int d^4x\,{\rm Tr}\sum_{n,m}\left[
    a_{\vec{n}}^*\alpha^\dagger_{\vec{n}} (i\partial_t+i\alpha_i\partial_i) b_{\vec{m}}\beta\alpha_{\vec{m}}\beta+
    b_{\vec{m}}^*\alpha^\dagger_{\vec{m}}\beta (i\partial_t+i\alpha_i\partial_i) a_{\vec{n}}\alpha_{\vec{n}}\beta\right]\nonumber\\
    =&\int d^4x\, \sum_{\vec{n},\vec{m}}{\rm Tr}\,\left[\alpha^\dagger_{\vec{n}}\alpha_{\vec{m}}\right]
    \Bigl(
    a_{\vec{n}}^* (i\partial_t) \Gamma_{\vec{m}} b_{\vec{m}}+
    b_{\vec{n}}^* (i\partial_t) \Gamma_{\vec{m}} a_{\vec{m}}\Bigr)\nonumber\\
    +&\int d^4x\,\sum_{\vec{n},\vec{m},i}{\rm Tr}\,\left[\alpha^\dagger_{\vec{n}}\alpha_i\alpha_{\vec{m}}\right]\Bigl(
    a_{\vec{n}}^* (i\partial_i) \Gamma_{\vec{m}} b_{\vec{m}}-
    b_{\vec{n}}^* (i\partial_i) \Gamma_{\vec{m}} a_{\vec{m}}\Bigr)
\end{align}
where $\beta\alpha_{\vec{p}}\beta=\Gamma_{\vec{p}}\alpha_{\vec{p}}$ and
$\Gamma_{\vec{p}}=\left(-1\right)^p=\pm 1$ according to whether the corresponding form is even
or odd.
This can be written more succinctly in terms of spatial K\"{a}hler-Dirac fields as
\begin{align}
    S&=\int d^4x\,\left[\overline{A}\,iK_s \Gamma\,B\right] -\left[\overline{B}\,iK_s\Gamma\,A\right]
    +\int d^4x\, \left[\overline{A}\,i\partial_t\, \Gamma B\right]+\left[\overline{B}\,i\partial_t\, \Gamma A\right]
\end{align}
where the subscript on the
K\"{a}hler operator $K=d-d^\dagger$ indicates that the exterior derivative is restricted to the
spatial manifold and the inner product now sums over spatial \KD fields. 
The action is real since $\Gamma$ anti-commutes with $K_s$ and manifestly invariant
under spatial diffeomorphisms. In fact it is actually invariant under spacetime diffeomorphisms which mix the $A$ and $B$ fields. However, as before the theory is not unitary which can be seen if one writes down
the canonical anti-commutators:
\begin{align}
    \{A(x),\bar{B}(y)\}&=\Gamma\delta^3(x-y)\nonumber\\
    \{B(x),\bar{A}(y)\}&=\Gamma\delta^3(x-y)
\end{align}
The factors of the spatial form parity $\Gamma$ appearing on the right hand side again again produce states with
negative norm. 
If we define 
\begin{align}
C(x)&=\frac{1}{\sqrt{2}}(A(x)+\Gamma B(x))\nonumber\\
D(x)&=\frac{1}{\sqrt{2}}(A(x)-\Gamma B(x))
\end{align}
we can diagonalize the action
\begin{equation}
    S=\int d^4x\,\left(\bar{C} iK_s C-\bar{D} iK_s D +\bar{C} i\partial_t C-\bar{D}i\partial_t D\right)
    \label{forms}
\end{equation}
The symmetries $\cal P$, $\cal T$ and $\cal C$ are given by
\begin{align}
    C(x)&\stackrel{{\cal P}}{\to} \Gamma C(x)\quad x\to -x\nonumber\\
    C(x)&\stackrel{{\cal T}}{\to} \Gamma C(x)\quad i\to -i\nonumber\\
    C(x)&\stackrel{{\cal C}}{\to} \bar{C}(x)
\end{align}
where $\Gamma$ is the spatial parity of the form which anticommutes with the spatial \KD operator. The transformations on the field $D(x)$ are the same.

The anti-commutation relations that follow from eqn.~\ref{forms} are
\begin{align}
    \{\bar{C}(x),C(y)\}&=\delta^3(x-y)\nonumber\\
    \{\bar{D}(x),D(y)\}&=-\delta^3(x-y)
\end{align}
Again, the presence of minus signs on the anti-commutator for D fields produces negative
norm states and we see that the theory is not unitary. Indeed $C$ and $D$ correspond to the matrix
fields $\Psi_+$ and $\Psi_-$ respectively.

As before, we can rectify this problem by introducing a modified inner product $\bra{\phi} J\ket{\psi}$
where $J$ anti-commutes with $D,\bar{D}$ and commutes with all $C$ field operators.
The norm of a single particle state is now
\begin{align}
    \bra{0}D_k|J|\bar{D}_k\ket{0}&=-\bra{0}JD_k\bar{D}_k\ket{0}=\bra{0}J\ket{0}\nonumber\\
    \bra{0}C_k|J|\bar{C}_k\ket{0}&= +\bra{0}JC_k\bar{C}_k\ket{0}=\bra{0}J\ket{0}
\end{align}
Explicitly it can be written 
\[J=e^{i\pi\int d^3k\; \overline{D}_kD_k}=\left(-1\right)^{N^{(-)}}\]
where $N^{(-)}$ counts the number of negative mode excitations $N^{(-)}=\int d^3k\, \bar{D}_kD_k$.
As noted before, the matrix elements of the $J$-deformed theory are no longer Lorentz invariant and indeed $D$ mixes
with $C$ under boosts.

\section{Summary and conclusions}

In Euclidean space the 
matrix field $\bar{\Psi}$ is generally taken to be independent of $\Psi$ and the theory in flat
space can be interpreted
as either a theory of four Dirac fermions with an $SU(4)$ flavor symmetry or equivalently as 
a theory of p-forms invariant under the diagonal subgroup of the Euclidean $SO(4)$ Lorentz symmetry and
an $SO(4)$ subgroup of the flavor symmetry.

In Minkowski spacetime one has two inequivalent
choices for the Wick rotation: $\bar{\Psi}=\Psi^\dagger \gamma^0$ - yielding a unitary, Lorentz invariant
theory of Dirac fermions or $\bar{\Psi}=\gamma^0\Psi^\dagger\gamma^0$ - giving rise to
a Lorentz invariant but non-unitary theory of integer spin forms. The non-unitary nature of the theory ia manifested by the appearance of negative norm states. We have
shown that the latter theory can
be rendered unitary by inserting an operator $J$ into the definition of the inner product in the space
which renders the latter positive definite. This operator anti-commutes with the modes giving rise to negative norm states and commutes with the Hamiltonian 
leading to unitary time evolution. However the price that is paid is that J does not commute with boosts and
the resultant theory is not Lorentz invariant as a theory of tensors.
One can understand this result as a consequence of the spin-statistics theorem which makes it impossible
to quantize an integer spin theory with anti-commutation relations while maintaining unitarity, locality, and Lorentz invariance
\cite{Pauli1940,Luders1958,Streater1964}. In a free theory at least the
net effect of the J-deformation is to remove all problematic minus signs so that the resultant J-deformed theory can be interpreted as a Lorentz invariant theory of four degenerate Dirac spinors once more.

Let us now reconcile our findings with what is known about anomalies. In \cite{Fidkowski_2023} it was proved that unitary fermionic lattice 
theories with a continuous $U(1)$ symmetry cannot manifest an anomaly.
In contrast in \cite{Catterall:2018lkj} it was shown that a certain onsite
$U(1)$ symmetry of massless Euclidean \KD fermions was broken to $Z_4$ when the fermions propagate on the sphere. The two claims are consistent 
when it is realized that the \KD theory is not reflection positive. Once one unitarizes
the theory by the insertion of $J$, the anomaly vanishes as required by the no-go theorem.
One simple way to
see this is to recognize that the partition function of the $J$-deformed and unitary Euclidean
theory should be computed
on a manifold with the topology ${\cal M}_3\times S^1$ which has vanishing Euler number. 

The appearance of an anomaly for Euclidean \KD fermions
is reminiscent of the anomaly reported for the two dimensional ghost system studied in \cite{Chang_2021}. Indeed in that case the anomaly can be computed
by inflow from a Wen-Zee topological term  \cite{WenZee1992}. In the case of \KD fermions 
the anomaly for the $U(1)$ symmetry can be obtained via an inflow argument
from a gravitational Chern Simons theory 
\cite{Catterall:2022jkn}. The remaining $Z_4$ symmetry of the Euclidean \KD theory
suffers from a mod 2 anomaly that can be canceled with two copies of the system.  It is intriguing that this field content originating in a non-reflection positive theory,
when expressed in terms of spinors in flat space, nevertheless 
corresponds to the sixteen flavors of chiral
fermion needed
to cancel the spin-$Z_4$ anomaly in Minkowski space. It appears that this connection deserves further scrutiny.

To summarize, it seems that the only unitary and Lorentz invariant formulation of \KD fermions
necessarily treats them as Dirac spinors. If one wants to keep the tensor
form that is natural in Euclidean space one must give up Lorentz invariance.

\acknowledgments
This work was supported in part by the U.S. Department of Energy (DOE) under Award Number DE-SC0009998.
S.C and J.H would like to acknowledge useful conversations with Shu-Heng Shao, Ho tat Lam, Tom Hartman and Alex Maloney.

\bibliographystyle{JHEP}
\bibliography{references}

\appendix

\section{Pseudo-hermiticity and positivity}
The modified inner product corresponding to the insertion of $J$ implies a modified notion
of the adjoint operator. We require
\[\braket{\phi|O|\psi}_J^*=\bra{\phi}JO\ket{\psi}^*=\bra{\psi}O^\dagger J \ket{\phi}=\bra{\psi}J(JO^\dagger J)\ket{\phi}\]
showing that the new adjoint $O^\dagger_J=JO^\dagger J$.
If the system's Hamiltonian satisfies   
\begin{equation}JH^\dagger J=H^\dagger_J=H\end{equation}
it is said to be pseudo-hermitian \cite{Mostafazadeh:2010jpa}.
With respect to the $J$-norm the time evolution is then seen to be unitary
\begin{equation}
    \braket{\phi(t)|J|\psi(t)}=\braket{\phi(0)e^{iH^\dagger t}Je^{-iHt}\psi(0)}=\braket{\phi(0)Je^{iHt}J^2e^{-iHt}\psi(0)}=\braket{\phi(0)|J|\psi(0)}
\end{equation}
Furthermore, 
using this property one can show the eigenvalues of $H$ are real for energy
eigenstates with non-zero J-norm.
\begin{equation}
    \braket{\psi_E|(JH-H^\dagger J)|\psi_E}=(E-E^*)\braket{\psi_E|J|\psi_E}=0
\end{equation}
These properties of very reminiscent of systems with PT-symmetry \cite{Bender:1998gh,Bender:2002vv,Bender:2007nj}. This should not be surprising since theorems exist that show that any pseudo-hermitian
system must possess an anti-linear symmetry.
The argument is simple. Let's assume
I have an anti-linear symmetry $A$ with $[A,H]=0$ 
and the eigenstates of $A$ are the same as those of $H$. Thus
$A\ket{\phi}=\lambda_A\ket{\phi}$ and $H\ket{\phi}=E\ket{\phi}$ then
\begin{equation}
    [A,H]\ket{\phi}=(E^*-E)\lambda_A\ket{\phi}=0\quad\rightarrow {\it E}\; {\rm is\; real}
\end{equation}
In this case the anti-linear symmetry corresponds to time reversal $t\to -t$ and $i\to -i$.~\footnote{In general it is possible
that $[A,H]=0$ but the two operators do not share a common set of eigenstates. In this case one can
only show that eigenvalues come in complex conjugate pairs.}

\section{Constructing a Euclidean path integral for the toy model}
To build a path integral we can use coherent states. In our simple toy model 
the coherent state  is defined by 
\begin{equation}
    \ket{\psi}=e^{-\psi\Psi^\dagger}\ket{0}=\ket{0}-\psi\ket{1}
\end{equation}
It is an eigenstate of $\Psi$:
\begin{equation}
    \Psi\ket{\psi}=\psi\ket{\psi}
\end{equation}
The following argument verifies this
\begin{align}
    \Psi\ket{\psi}&=\Psi\ket{0}-\Psi\psi\ket{1}\nonumber\\
    &=0+\psi\Psi\ket{1}\nonumber\\
    &=\psi\ket{0}\nonumber\\
    &=\psi(\ket{0}-\psi\ket{1})\nonumber\\
    &=\psi\ket{\psi}
\end{align}
Similarly we can define an adjoint coherent state as
\begin{equation}
    \bra{\psib}=\bra{0}e^{-\Psi\psib}=\bra{0}-\bra{1}\psib
\end{equation}
satisfying
\begin{equation}
\bra{\psib}\Psi^\dagger=\bra{\psib}\psib\end{equation}
The inner product of such states is
\begin{align}
    \braket{\psib |\psi}&=(\bra{0}-\bra{1}\psib)(\ket{0}-\psi\ket{1})\nonumber\\
    &=\braket{0|0}+\bra{1}\psib\psi\ket{1}\nonumber\\
    &=1+\psib\psi\nonumber\\
    &=e^{\psib\psi}
\end{align}
Thus the correctly normalized coherent state is
\begin{equation}
    e^{-\frac{1}{2}\psib\psi}e^{\psi\Psi^\dagger}\ket{0}
\end{equation}
The corresponding resolution of the identity using coherent states is then given
in terms of the Grassmann integral.
\begin{equation}
    I=\int \ket{\psi}\bra{\psib}\left[\braket{\psib|\psi}\right]^{-1}d\psib\,d\psi=\int\ket{\psi}\bra{\psib}e^{-\psib\psi}d\psib\,d\psi
\end{equation}
One can verify this identity easily
\begin{align}
    \int \ket{\psi}\bra{\psib}e^{-\psib\psi}d\psib\,d\psi&=\int \ket{\psi}\bra{\psib}(1-\psib\psi)d\psib\,d\psi\nonumber\\
    &=\int (\ket{0}-\psi\ket{1})(\bra{0}-\bra{1}\psib)(1-\psib\psi)d\psib\,d\psi\nonumber\\
    &=\ket{0}\bra{0}+\ket{1}\bra{1}=I
\end{align}
Notice how this depends on the inner product of the states.
The partition
function of this system is given by
\begin{equation}
    Z={\rm Tr}\,e^{-\beta H(\Psi^\dagger, \Psi)}
\end{equation}
where the  Hamiltonian is just $H=\Psi^\dagger\Psi$.
To construct a Euclidean path integral representation for $Z$ one needs to split the time
interval $\beta$ into a series of elementary steps $\epsilon=\frac{\beta}{N}$
\begin{equation}
    e^{-\beta H}=\lim_{N\to\infty} (e^{-\epsilon H})^N
\end{equation}
Introducing the identity operator $N-1$ times and equating the initial
and final states to get a trace we find
\begin{align}
    Z&=\int \prod_{i=N-1}^0\bra{\psib_{i+1}}e^{-\epsilon H(\Psi^\dagger, \Psi)}\ket{\psi_{i}}\,
    e^{-\psib_{i}\psi_{i}}d\psib_i\,d\psi_i\nonumber\\
    &=\int \prod_{i=N-1}^{0} e^{\psib_{i+1}\psi_i}e^{-\epsilon H(\psib_{i+1},\psi_{i})}e^{-\psib_{i}\psi_{i}}d\psib_i\,d\psi_i\nonumber\\
    &=\int \prod_{i=0}^{N-1} \exp{\left(\left[\frac{1}{\epsilon}(\psib_{i+1}-\psib_i)\psi_i-H(\psib_{i},\psi_i)\right]\epsilon\right)}\,d\psib_i\,d\psi_i
\end{align}
where the quantum operators residing in $H$ have been replaced by
their (Grassman) valued eigenvalues in the second line and
we have replaced $\psib_{i+1}$ by $\psib_i$ in the Hamiltonian in the last line which is good to first order in $\epsilon$.
In the limit $N\to\infty$ and hence $\epsilon\to 0$ we find
\begin{equation}
    Z=\int D\psib\,D\psi\vert_{\rm APBC} \, e^{S(\psib,\psi)}
\end{equation}
where 
\begin{equation}S=\int_0^\beta \left[-\psib(\tau)\frac{d}{d\tau}\psi(\tau)-H(\psib(\tau),\psi(\tau))\right] d\tau\end{equation} with antiperiodic boundary conditions on the fermions.
This is the Euclidean path integral representation of $Z$. It should be clear that such a construction is only possible because of the existence of a well-defined resolution of the identity which requires all
states to be positive norm.

Let us see what happens in the case of a non-unitary system.
In this case the inner 
product $\braket{\psib|\psi}\to e^{-\psib\psi}$ with the result that the expression for the
identity operator becomes
\begin{align}
    I&=\int \ket{\psi}\bra{\psib}e^{\psib\psi}d\psib\,d\psi\nonumber\\
    &=-\ket{0}\bra{0}+\ket{1}\bra{1}
\end{align}
Thus this theory does not possess a bona fide
identity operator. The absence of a well defined
identity prevents us from constructing a Euclidean path integral along the lines described earlier.

Again the situation can be rectified if one replaces the usual inner product by its J-modified cousin
$\braket{\psib|\psi}_J$ corresponding to the Grassmann integral:
\begin{equation}
    I=\int\ket{\psi}\bra{\psib}e^{\psib\psi}J\,d\psib\,d\psi
\end{equation}
The resulting partition function then also involves an insertion of $J$:
\begin{equation}
    Z={\rm Tr}\left[J\,e^{-\beta H}\right]={\rm Tr}\left[\left(-1\right)^{N}\,e^{-\beta H}\right]
\end{equation}
where $J=\left(-1\right)^N$ with $N=\Psi^\dagger\Psi$ the fermion number. 
\end{document}